# "Exclusive-OR" operation with Fano resonant MEMS Metamaterial

**Manukumara Manjappa[1,2], Prakash Pitchappa[1,2], Navab Singh[5], Nan Wang[5], Nikolay I Zheludev[2,6], Chengkuo Lee[3,4], Ranjan Singh[1,2,*]**

[1]*Division of Physics and Applied Physics, School of Physical and Mathematical Sciences, Nanyang Technological University, 21 Nanyang Link, Singapore 637371, Singapore.*

[2]*Centre for Disruptive Photonic Technologies, The Photonics Institute, 50 Nanyang Avenue, Nanyang Technological University, Singapore 639798.*

[3]*Department of Electrical & Computer Engineering, National University of Singapore, 4 Engineering Drive 3, 117576, Singapore.*

[4]*NUS Suzhou Research Institute (NUSRI), Suzhou, Industrial Park, Suzhou, P. R. China 215123.*

[5]*Institute of Microelectronics, 11 Science Park Road, 117685, Singapore.*

[6]*Optoelectronics Research Centre and Centre for Photonic Metamaterials, University of Southampton, Highfield, Southampton SO17 1BJ, United Kingdom.*

\* Corresponding author: *ranjans@ntu.edu.sg.*

In recent years, a range of reconfigurable metamaterials controlled with thermal, electric, magnetic and optical signals have been developed for dynamically manipulating the intensity, phase and wave-front of electromagnetic radiation across the broad electromagnetic spectrum ranging from microwave to optics. Here for the first time, we demonstrate a reconfigurable metasurface performing "exclusive OR" (XOR) logical operation using two independent control electrical inputs and an optical read-out in the form of far-field Fano intensity states at terahertz wavelengths. At the same time, the near field resonant confinement enables a universal NAND logical operation. Further, by using a single electrical input control and an optical readout, a logical NOT operation is achieved at the far-field intensity states of the Fano resonance. The proposed reconfigurable microcantilever based Fano design that exhibits multiple logical operations can create a versatile platform for realization of programmable and randomly accessible metamaterials with enhanced electro-optical performance, multichannel data processing and encryption techniques for high speed wireless networks which are now being pushed towards terahertz wavelengths.

Recent trend in the metamaterial research is greatly advancing towards realizing functional and reconfigurable metamaterials[1-3] that enable a real-time control over their structural and optical properties, thereby creating exceptional opportunities in the field of active metamaterials and plasmonics. They possess unique advantage in active manipulation of the near fields in all the three spatial directions by exploiting sensitive changes to their micro/nano scale movements, which makes them more resilient and merits their applications as future generation state-of-the-art active photonic devices. The dynamic control over the near-field features has provided a new paradigm of controlling and manipulating the light-matter interactions, which gives enhanced and tunable nonlinear properties in the metamaterial. Over the years, a range of tunable metamaterials have been demonstrated, where their optical features are actively controlled through their structural reconfiguration using the electrical controls[4-7], magnetic fields[8,9], thermal gradients[10-12] and optical pulses[13-15]. A class of reconfigurable metamaterials based on micro/nano-electro-mechanical systems (MEMS/NEMS) mainly operating in the terahertz (THz) to near-infrared frequencies have enabled dynamic manipulation of THz properties such as magnetic response[4,10], transparency[16], near-perfect absorption[17] and THz anisotropy[5]. However, the ability to flexibly control near- and far-field interactions by establishing independent structural reconfigurations of meta-atoms at the unit cell regime of metamaterial has been challenging.

The future prospects and challenges of functional metamaterials lie in achieving multiple controls within the unit cell of the metamaterial, which could provide a flexible platform for realizing extremely versatile metamaterials manifesting enhanced electro-optical performance[18]. Multiple controls within the unit-cell will enable a precise tailoring of near-field interactions between the meta-atoms by relatively maneuvering their structural properties, thereby obtaining the optical properties on demand. The nano/micro sized switchable metamaterials[2,7] have given a promising pathway to control near- field coupling between the metamolecules by establishing the independent/multiple controls over the structural reconfiguration of the constituent resonators within the unit cell of the metamaterials. However, such demonstrations have been solely restrained to investigating the system responses in terms of single-input-output (SIO) configurations. In the hindsight of enhancing their multifunctional capabilities in digital[19-21] and multiple signal processing applications, one way is to establish multi-valued dependency between the input and output characteristics of the metamaterial. Here, we experimentally realize excitation of sharp Fano resonance in a MEMS reconfigurable metamaterial exhibiting multiple-input-output (MIO) characteristics in its near- and far-field optical properties. By preparing the structural geometry of the MEMS-Fano metamaterial device in various meta-stable states, we realize "exclusive OR" (XOR) and NAND logical operations in the far-field and the near-field optical characteristics, respectively by using two independently controlled electrical inputs and an optical/near field read-out at terahertz wavelengths. XOR is a very important function that is not linearly separable into more basic logical functions, which makes it more resilient and useful in the information and computational technologies as parity generators, one-time pad (OTP) based unbreakable cryptography protocols[22,23], pseudorandom number generators and digital encoders or decoders in signal processing. It serves as a key component in the central processing unit (CPU), performing full- and half-adder operations in storing and processing the data. On the other hand, the NAND logical operation is a functionally complete set logical operator which can be used to express all the basic set logical operations by defining a network of NAND gates. Further, we show that the SIO configuration of Fano excitation can constitute the NOT logical operation using the single electrical control and optical readout at the far-field intensity state of Fano resonance. The multi-logical operations of the proposed MEMS Fano design together with the volatile and nonvolatile[24,25] regimes of MEMS actuation can enhance the digital functionalities of the metamaterials in realizing optical memory registers to encode, harvest, process and send the information in form of encoded/decoded optical bits across the far- to near-infrared frequencies.

**Design and Fabrication**

To precisely elucidate the operation of XOR and NAND functionalities possessing MIO features in the metamaterial, we fabricated a micro-electro-mechanical systems (MEMS) based metamaterial design consisting of two split ring resonators (SRR-1 and SRR-2) that are independently actuated by applying the voltages $V_1$ and $V_2$, respectively (shown in Fig. 1(a)). The device is fabricated using the photolithography technique, where periodic array of SRRs made of deformable bimorph structures consisting of 900 nm thick aluminum (Al) on top of 50 nm aluminum oxide ($Al_2O_3$) layer are patterned on a lightly doped silicon (Si) substrate (refer to Methods Section for the device fabrication details and characterization, and the Supplementary Information Fig. S1 - S4). Due to the residual stress in the bimorph layers, the cantilevers are bent up, thereby increasing their released heights (*h*). Scanning electron microscope image of the fabricated MEMS Fano-metamaterial is shown in Fig. 1(a) in the colored scale that illustrates the maximum asymmetric state of the device with SRR-1 snapped down on the substrate using voltage $V_1$ = 30 V and SRR-2 retained in the released state of the bimorph cantilevers with $V_2 = 0$ V. The out-of-plane reconfiguration of the released cantilevers is achieved through electrostatic actuation, by applying voltage across the Al layer and silicon substrate. Metal lines connecting the SRR-1 and SRR-2 cantilevers are electrically isolated from each other and this allows for the independent reconfiguration of $h_1$ and $h_2$ through application of voltage $V_1$ and $V_2$, respectively. The selective reconfiguration at the sub-unit cell level provides the flexibility to introduce dynamically tunable structural asymmetry along the $z$-axis, defined by an effective asymmetry parameter, $\delta = \left|\frac{h_1 - h_2}{s}\right| \times 100\%$, where ($h_1$, $h_2$) and $s$ are the released heights and length of the cantilever arms of SRR-1 and SRR-2, respectively. Fig. 1(b-d) are the SEM images of the unit cell showing the sequential control of SRR cantilevers by applying voltages $V_1$ and $V_2$ across the Al-metal lines and the silicon substrate.

**Results**

**<u>Active control of Fano resonances:</u>** Persistent control of resonance features in the MEMS metamaterial is experimentally characterized by using the photoconductive antenna based THz-TDS setup in the transmission mode (ref. to the Methods Section for more details). The measured transmission spectra for the increasing voltage ($V_1$) are shown in Fig. 2(a), where the inset diagram presents the experimentally measured electrostatic actuation profile of the cantilevers by applying the voltage on one of the SRRs. Initially, for the case where no voltage is applied across the resonators i.e. $V_{1,2} = 0$ (see, Fig. 1(b)), the arms of the two SRRs are symmetrically inclined at same heights $h$ ($\delta = 0$) along the $z$-direction that results in the excitation of strong dipole type of resonance at 0.78 THz for the incident THz radiation polarized in the $E_y$ direction. When voltage ($V_1$) is applied across the Al lines of the released cantilevers (say, SRR-1) and Si substrate, the SRR-1 cantilevers gradually deflect towards the fixed substrate due to the attractive electrostatic force. This deformation in the height of the SRR-1 cantilevers creates a structural asymmetry ($\delta$) along the $z$-axis of the sample. As a result, near field coupling between the asymmetric structures exhibits Fano-type of interference effects that excites a sharp trapped mode (Fano-type) resonance[26,27] feature (at 0.6 THz) within a broad dipolar resonance. Upon continuously increasing $V_1$ across the SRR-1, strength of Fano resonance grows and reaches its maximum amplitude for $V_1 = 30$V (where $V_2 = 0$V), as shown in Fig. 2(a). Subsequently, when the voltage $V_2$ is applied across SRR-2, its cantilever arms are gradually pulled towards the substrate, as a result the asymmetry in the structure progressively decreases. Due to the decreasing asymmetry, the Fano resonance starts to weaken and shows a monotonic decrease with increasing $V_2$ and completely diminishes as $V_2$ finally reaches 30 V, as shown in Fig. 2(b) (both SRR-1 and SRR-2 are snapped down on the substrate with $V_{1,2}$ = 30 V (see, Fig. 1(d))) and symmetry is restored in the system.

The correspondence between the transmission spectra obtained by the sequentially applied voltages ($V_1$ and $V_2$) in the experiments and the structural asymmetry ($\delta$) in the MEMS metamaterial is established by the numerical simulations. The transmission spectra for varying $\delta$ are obtained using finite difference time domain (FDTD) calculations offered by commercially available CST microwave studio software, as shown in Fig. 2(c) and (d). The value of the structural asymmetry parameter ($\delta$) is estimated using the expression for $\delta$ based on the experimentally measured inclined heights ($h_1, h_2$) of the cantilevers of resonators SRR-1 and SRR-2 (as shown in inset Fig. 2(a)). The insets in Fig 2(c) and (d) represent the sequential actuation of SRR-1 and SRR-2 resonators that corresponds to the continuous increase and decrease in the structural asymmetry parameter, which signifies one complete ramp of both asymmetry parameter ($\delta$) and the sequentially applied voltages $V_1$ and $V_2$. As a first actuation sequence, the asymmetry is increased by decreasing the released height ($h_1$) of SRR-1, which results in the strengthening of the Fano resonance feature that reaches its largest resonance amplitude at the maximum asymmetry of $\delta$ = 2.3 %. In the next actuation sequence, upon decreasing the released height ($h_2$) of SRR-2, the amplitude of the Fano resonance diminishes and eventually disappears as the cantilever of SRR-2 touches down on the substrate ($\delta$ = 0).

**Multiple-input-output characteristics**: The most striking feature of the MEMS Fano-metamaterial is the observed anisotropic nature in the excitation of Fano resonances. The anisotropic Fano coupling exhibits two distinctive pathways (output states) for the transmitted Fano intensity ($|\Delta T|$) and the near-field characteristics with respect to the sequential application of two voltage inputs ($V_1$ and $V_2$) or the increasing and decreasing pathways of the structural asymmetry parameter ($\delta$) (Detailed plots on *Q*-factors and Figure of Merits are given in the Supplementary Information). The calculation and definition of transmission intensity Fano resonance $|\Delta T|$ is discussed in the Supplementary Information figure S5. In Fig. 3(a), $|\Delta T|$ of Fano resonance obtained from measurements is plotted against the applied voltages $V_1$ and $V_2$, where the red circles and the green squares represent increasing and decreasing Fano intensity with the sequential control of $V_1$ and $V_2$, respectively. Although this scenario showing the distinctive pathways resembles the hysteresis behavior as observed in many natural phase change materials ($VO_2$ and Ferrites), but indeed it is a two output states for two input controls ($V_1$ and $V_2$) to form a closed loop in $|\Delta T|$. This behavior in the optical response of the system constitutes multiple-input-output (MIO) configuration in the intensity state of the Fano resonance. In the numerical simulations, the sequential control of the two voltage inputs ($V_1$ and $V_2$) is expressed as one control parameter in terms of asymmetry ($\delta$) of the structure, which is a critical parameter for the excitation of Fano resonances. In Fig. 3(b), the transmission intensity ($|\Delta T|$) of Fano resonance is plotted for the increasing and decreasing scenario of structural asymmetry parameter ($\delta$), where the SRR-1 and SRR-2 are actuated sequentially (as shown in the insets of Fig. 3(b)). The output intensity response in the far field displays two distinct values for the same asymmetry parameter ($\delta$) depending on the increasing or decreasing asymmetry sequence in the system, mimicking a hysteretic type of closed loop.

**<u>Exclusive OR metamaterial</u>**: The uniqueness of digitizing the excitation of Fano resonance in terms of its far-field intensity states by preparing the microcantilevers of SRRs in various structural metastable (actuation) states using two electrical controls in the proposed MEMS Fano-MM constitutes a digital exclusive OR (XOR) logical function. The schematic of the THz characterized MEMS Fano-metamaterial based XOR logical operation is described in Fig. 3(c). The structural states ('up' or 'down') of the constituent resonators SRR-1 and SRR-2 are independently reconfigured using the voltage inputs $V_1$ and $V_2$, respectively in determining the output state of Fano resonance (*F*). The structural metastable states of the resonators determined by the electrical inputs ($V_{1,2}$) are represented by the logical binary digits, where 'up-state' of the resonator corresponds to binary '0' ($V_{1,2} = 0$ V) and the 'down-state' corresponds to binary '1' ($V_{1,2} = 30$ V). The true (ON) and false (OFF) states of the Fano resonance intensity in the far-field are represented by the binary digits '1' and '0', respectively. The measured THz far-field spectra for the various metastable states of the MEMS resonators are given in Fig. 3(d) (i-iv). The voltage inputs ($V_1$ and $V_2$) applied to the individual resonators (SRR-1 and SRR-2) are programmed using sequential trigger bits {0,1} that controls the actuation heights (up/down) of SRR-1 and SRR-2, respectively. Since, the Fano resonance feature results due to the asymmetry in the structural configuration of the metamaterial, for input voltages ($V_1 = 0$, $V_2 = 30$ V and $V_1 = 30$ V, $V_2 = 0$ V) there exists two asymmetric structural configurations 'up-down' (0,1) and 'down-up' (1,0) that results in the true state for the Fano resonance condition (i.e. *F* = 1), as shown in Fig. 3(d) (ii) and (iii), respectively. On the other hand, for the symmetric structural configurations of the device ($V_1 = 0$, $V_2 = 0$ V and $V_1 = 30$ V, $V_2 = 30$ V), 'up-up' (0,0) and 'down-down' (1,1) results in the absence of Fano resonance ($F = 0$) state, as shown in Fig. 3(d) (i) and (iv) respectively. The resulting truth table is presented in the inset of Fig. 3 (c), which resembles the digital XOR operation, where the Fano output is false ($F = 0$) if and only if both the inputs are true (1) or false (0), otherwise, Fano output shows the true ($F = 1$) output state. One of the important aspects of MEMS based metamaterial device is that it can be operated in both volatile as well as in nonvolatile regimes depending on the partial or complete actuation of the constituent resonators' cantilevers in the metamaterial, respectively. This directly reflects on the volatile and nonvolatile operation of the demonstrated XOR functionality in MEMS Fano-metamaterial. The results discussed in Fig. 3(d) (i-iv) represents non-volatile operation regime of the MEMS Fano-metamaterial, where due to the stiction in the MEMS devices, microcantilevers will stay in contact with the substrate even after the input voltage is removed. This nonvolatile property of the device may affect the speed and repeatability of the device operation, but enables the memory features in the device, which could be potentially used as memory registers in data storing and processing techniques. On the other hand, the volatile feature of MEMS Fano-metamaterial enacted XOR operation enables the partial actuation of SRR-1 and SRR-2 cantilevers for the maximum applied voltage of $V_{1,2} < 30$ V assures persistent repeatability of the device operation. Thus, by independently controlling the voltage states ($V_{1,2}$) of SRR-1 and SRR-2 using '0' (OFF) state and '1' (ON) state, the Fano excitation based logical switch is shown to exhibit both volatile and nonvolatile regime of XOR logical operation.

**<u>NOT logical operation</u>**: The proposed design provides a flexibility of tuning the asymmetry of the structure to show either anisotropic or isotropic coupling between the adjacent meta-atoms just by adequately coding the input electrical signaling sequence. Fig. 4(a) represents the experimentally measured variation in the intensity of the Fano resonance with respect to the voltage $V_2$ applied on SRR-2, by keeping SRR-1 in contact with the substrate with $V_1 = 30$ V (decreasing pathway of the asymmetry). For the voltage $V_2 = 0$ ($V_1 = 30$ V), the Fano resonance possesses the maximum intensity ($\Delta T$) value and by increasing voltage ($V_2$), $\Delta T$ gradually decreases and finally becomes zero at $V_2 = 30$ V, as the symmetry is restored in the system. In the initial structural state of the device, where the SRR-1 is put in contact with the substrate and the SRR-2 is kept in the released state ($V_2 = $ '0') shows the asymmetric structural configuration of the device that results in the excitation of the Fano resonance in the far-field intensity state ($F = 1$). As the voltage $V_2 = 30$ V (i.e. $V_2 = $ '1') is applied, the microcantilever arms of SRR-2 resonator comes in contact with the substrate thereby restoring the structural symmetry of the device that annihilates the Fano resonance feature in the structure (i.e. $F = 0$). This correlation between the input voltage states of SRR-2 and the optical readout in the form of Fano resonance far-field intensity state manifests the logical NOT operation, as shown by the truth table presented in inset Fig. 4(a). Further, numerically calculated ΔT values plotted in Fig. 4(b) shows isotropic variation with respect to decreasing and increasing pathways of the asymmetry (δ), where by keeping the SRR-1 in contact with the substrate, the released height of SRR-2 is decreased and increased respectively, as shown in the inset of Fig. 4(b). This configuration of Fano tuning signifies the single input-output (SIO) characteristics in the MEMS Fano-metamaterial, which is due to the isotropic nature of coupling between the resonators during the increasing and decreasing pathways of asymmetry.

**NAND logical operation:** In addition to the XOR and NOT logical operations using the far-field characteristics, the near-field characteristics reveals the universal NAND logical operation in the form of confined electric fields measured at the tip of the microcantilever resonators in their ON and OFF states. Numerically calculated electric field strengths are plotted in Fig. 5(a), where the absolute E-field for structural configurations (i), (ii) and (iii) shows enhanced field strengths compared to structural configuration (iv), where both microcantilevers are prepared in ON states. The absolute strengths of the confined electric fields are plotted in Fig. 5(b) that highlights the distinctive variation for the increasing and decreasing pathways of the asymmetry. For the symmetric state with both the microcantilevers prepared in the OFF states (i), the electric field confinement is nearly an order of magnitude greater than the symmetric configuration with both the microcantilevers are prepared in the ON state (iv). Whereas, for the two asymmetric configurations of the microcantilevers (ii and iii), the field confinement in the structure shows similar strengths to the symmetrically prepared OFF state (i) of the structure. Higher strength of the electric field confinement is labeled as the binary '1', whereas lower electric field strength is represented by binary '0', as shown in Fig. 5(b). Therefore, the change in the confined electric field strength measured at the tip of the microcantilevers in their various metastable structural configurations (i - iv) constitutes a logical NAND function as described by the truth table given in Fig. 5(c). The NAND operation signifies that near field output state of the device is true (logical '1') when either one or both the input structural states of the device are false (logical '0'), otherwise it results in the false output state (logical '0'). The NAND logical operation is significant owing to its unique feature of functional completeness (universal gate), as any boolean function can be implemented by using the combination of several NAND gates.

**Discussion**

The flexibility to operate the device with the MIO and the SIO characteristics using the two independent input controls provides the useful feature of realizing multiple logical operations in its near-field and the far-field optical characteristics. The observed MIO optical characteristics in the proposed MEMS Fano-metamaterial exploits the nonlinear spatial distribution of near-fields extended along the *z*-axis of the sample. Independent and sequential actuation of SRR-1 and SRR-2 cantilevers mediates the anisotropic Fano coupling between the resonators that show contrasting coupling strengths depending on whether the adjacent fixed resonator cantilever is closer or farther away from the surface of the substrate. Continuous and sequential actuation of the adjacent resonators plays a critical role in the observed anisotropic excitation of Fano resonances that illustrates the MIO states for a given $V_1$ and $V_2$ (or $\delta$). As seen from inset of Fig. 3(b), during the increasing $\delta$ configuration, as the fixed resonator cantilever (SRR-2) is positioned away from the substrate, hence the resonator arm is in the vicinity of a weak spatial field distribution. Therefore, it requires larger structural asymmetries (higher excitation threshold) to excite the Fano resonance. While decreasing the asymmetry, the SRR-1 positioned on the substrate is likely to possess greater influence on the near field coupling occurring in the asymmetric structure that results in stronger excitation of Fano features at smaller asymmetries (lower excitation threshold).

In summary, we demonstrate excitation of sharp Fano resonances in a MEMS based metamaterial using two independent voltage controls that constitutes a digital XOR, NOT and NAND logical gates in its far- and near-field optical properties in the terahertz wavelengths. Generally, the XOR functionality possesses unique feature of pseudorandom generation and serves as a key component in one-time pad encryption/decryption techniques in establishing a theoretically secured cryptographic protocols. The NAND logical operation being the universal logical function would benefit to construct all the other boolean logical operation, thereby providing a flexibility of enhancing the digital functionalities of the device. Our results can show potential prospects in super-encryption techniques in *i*-banking sectors, short messaging services (SMS), defense, national data security systems and high speed wireless communication networks, which are now being pushed towards terahertz wavelengths[28,29]. The reported multi-functionalities of the proposed MEMS Fano-metamaterial are largely suitable for real world applications such as active sensors possessing tunable mode volumes, tunable nonlinear devices and modulators. Alongside, the MIO characteristics of the MEMS Fano-metamaterial could potentially provide a flexible platform for developing the next generation randomly accessible, digital and programmable metamaterials for precise tailoring of optical properties and multichannel data processing across the wide band of electromagnetic spectrum.

**Methods**

**Sample Fabrication**: The MEMS Fano-metamaterial was fabricated using a complementary metal-oxide-semiconductor (CMOS) compatible process as described below. First, the lightly doped 8 inch silicon substrate of 725 μm thickness was cleaned and a 100 nm thick sacrificial $SiO_2$ layer was deposited using low pressure chemical vapor deposition (LPCVD) process. Following this conventional photolithography process was used to pattern the anchor region. With the designed pattern, the parts of sacrificial $SiO_2$ for anchor regions were dry etched using reactive ion etching process. After this, a 50 nm thick $Al_2O_3$ layer was deposited using the ALD process, followed by the sputter deposition of 900 nm thick Al. Note that the bimorph layers ($Al/Al_2O_3$) were in physical contact with Si substrate at the anchor region, and in the remaining part of the wafer, it was on top of sacrificial $SiO_2$ layer. Then, the second photolithography step was carried out for defining the cantilevers and metal lines of metamaterial patterns. Following this, both Al and $Al_2O_3$ layers were dry etched to form the designed metamaterial. Finally, vapor hydrofluoric was acid (VHF) was used to isotropically etch the $SiO_2$ sacrificial layer underneath the bimorph structures, thereby suspending it over the Si substrate with an air gap between them. At the anchor region, since the bimorphs were in physical contact with Si substrate; the VHF release process was not time controlled, and this ensured higher yield of the devices. Due to the residual stress in the bimorph cantilevers, the released cantilevers were bent up, thereby increasing the initial tip displacement.

**Electromechanical Characterization of the MEMS device:** The deflection/actuation profiles of released microcantilevers were measured using Lyncee Tec. reflection digital holographic microscope (R-DHM). The released chips are wire bonded to a printed circuit board (PCB). Separate voltage supplies ($V_1$ and $V_2$) are used for the actuation of SRR 1 and SRR 2 microcantilevers, respectively. Silicon (Si) substrate was chose as the ground potential, and the microcantilevers were positively biased. When voltage is applied across the released cantilevers and Si substrate, the attractive electrostatic force deforms the suspended cantilevers towards the fixed Si substrate. This mechanical deformation of cantilevers induces a restoring force that opposes the electrostatic force causing the deflection at the first place. Hence, the final position of the cantilever at a given voltage is determined by the equilibrium position, where the electrostatic force and restoring force balances each other. As the applied voltage increases, the electrostatic force increases much higher than the restoring force and at a critical value known as the Pull in voltage (> 25 V), the electrostatic force will be higher than the restoring force, thereby bringing the microcantilevers to be in physical contact with Si substrate (shown in inset of Fig. 2(a)).The pull-in can be clearly observed through the optical microscope fitted on the R-DHM. The $Al_2O_3$ layer beneath the Al layer, will ensure that there is no current flowing from Al layer to Si substrate, when pull in occurs. This is crucial because if the current flows through the Al/Si junction, then the temperature will rise up locally thereby melting the Al tips with the Si substrate, and will cause a permanent damage to the device.

**THz Measurements:** The MEMS Fano-metamaterial is optically characterized using the conventional GaAs photoconductive switch based THz-time domain spectroscopy system operating in the transmission mode. The wire bonded metamaterial sample is positioned at the focus of the THz beam. The electrical connections to the SRR-1 and SRR-2 resonators structures are established using a two channel DC voltage source. For four configurations of the voltages ($V_1 = 0$ V, $V_2 = 0$ V; $V_1 = 25$ V, $V_2 = 0$ V; $V_1 = 0$ V, $V_2 = 25$ V and $V_1 = 25$ V, $V_2 = 25$ V), the THz wave of beam spot 3 mm is incident in the normal angle on the sample and the transmitted THz pulse is captured using the THz detector connected to the lock-in amplifier. THz response through the bare silicon substrate is measured for the referencing the signal through the sample. In the post processing steps, the detected THz pulses measured through the sample and the bare substrate are fast Fourier transformed (FFT) to obtain the corresponding THz spectra. Later, the transmitted THz spectrum thorough the sample ($T_s(\omega)$) is normalized with respect to the transmission thorough the substrate ($T_R(\omega)$), i.e. $T_s(\omega) = T_s(\omega)/T_R(\omega)$ and the normalized spectrum is shown in the Figs. 2(a), 2(b) and Figs. 4(b-e).

**Numerical Simulations:** Finite-difference time-domain (FDTD) numerical simulations were conducted to calculate the THz transmission spectra and the confined electric near-fields and surface current distributions corresponding to the resonance modes for the normal incident of THz waves of TE polarization. Full-field electromagnetic wave simulations were performed using the commercial simulation software Computer Software Technology (CST) Microwave studio. For the material property, aluminum (Al) of thickness 900 nm was modeled as a lossy metal with conductivity of 3.57e7 S/m. Aluminum oxide and Silicon were modeled as lossless dielectric materials with dielectric constant of 9.5 and 11.9, respectively. In the simulation, a single unit cell of the metamaterial structures was simulated with periodic boundary conditions employed in axial directions orthogonal to the incident waves. The perfectly matched layers are applied along the propagation of the electromagnetic waves. Plane waves were incident into the unit cell from the port on the metal side, while the transmission spectrum was determined from the probe placed at the other side of metamaterial. The experimentally measured (inset Fig. 2(a)) deformation angles for the cantilevers are used to tilt the metal cantilevers that establishes the congruence between the values of applied voltages and the structural asymmetry used in the simulations. In the meanwhile, field monitors are used to collect the electric fields, magnetic fields and the respective surface currents at Fano resonance frequencies for varying asymmetry values.

**Figures**

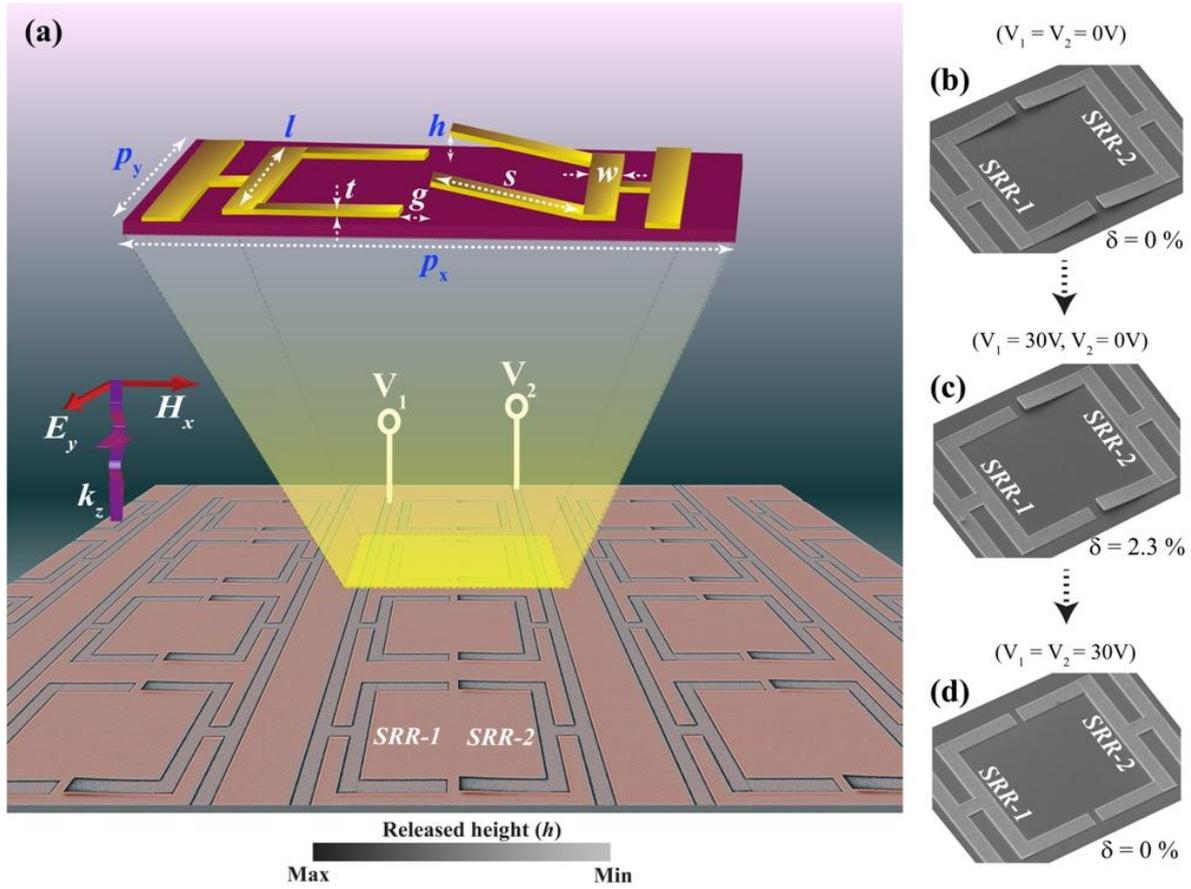

**Figure 1.** (**a**) Colored scanning electron microscope (SEM) image of the MEMS Fano-metamaterial. The unit cell comprises of two SRRs separated by a gap *g* and their cantilever arms of length *s* are released at a height *h*. The unit cell dimensions are depicted in the inset, where $p_x$ : 110 μm; $p_y$ : 75 μm; *l* : 60 μm ; *s* : 25 μm ; *w* : 6 μm ; *g* : 4 μm ; and *t* : 900 nm. $V_1$ and $V_2$ are the input voltage ports to achieve the independent actuation of SRR-1 and SRR-2 respectively. (**b-d**) SEM images of the unit cell showing the sequential actuation of SRRs with voltage $V_1$ and $V_2$ applied across the two SRRs, where the sequence from (**b**) to (**c**) represents the increasing asymmetry (δ) and (**c**) to (**d**) represents the decreasing asymmetry configuration.

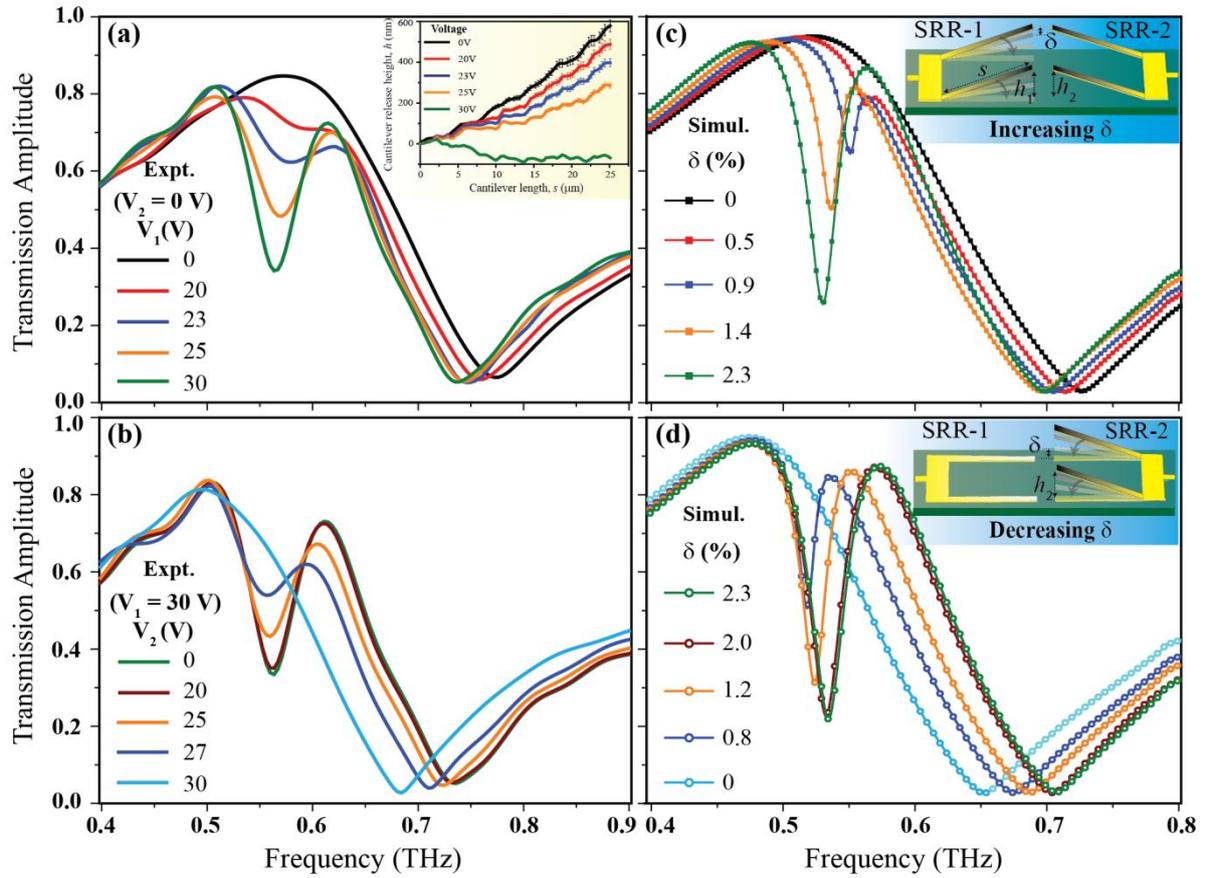

**Figure 2**. (**a**) Depicts the experimentally measured THz transmission spectra showing the evolution of Fano resonance for continuous actuation of SRR-1 by varying voltage $V_1$, while keeping $V_2 = 0$ V. Inset figure depicts the experimentally mapped actuation angles of the micro-cantilevers under the applied voltage (V) for the designed MEMS Fano-metamaterial. (**b**) Represents the measured spectra resulting from the actuation of SRR-2 by increasing $V_2$, while keeping $V_1 = 30$ V. (**c**) and (**d**) numerically simulated THz transmission spectra for increasing and decreasing structural asymmetry configurations of the proposed Fano MEMS metadevice. The values of the depicted asymmetry parameter ($\delta$) in (c) and (d) show one-to-one correspondence with the voltage values of $V_1$ and $V_2$ varied in (a) and (b), respectively. The insets signify the sequential actuation of SRR-1 and SRR-2 respectively, showing the increasing and decreasing structural asymmetry configurations.

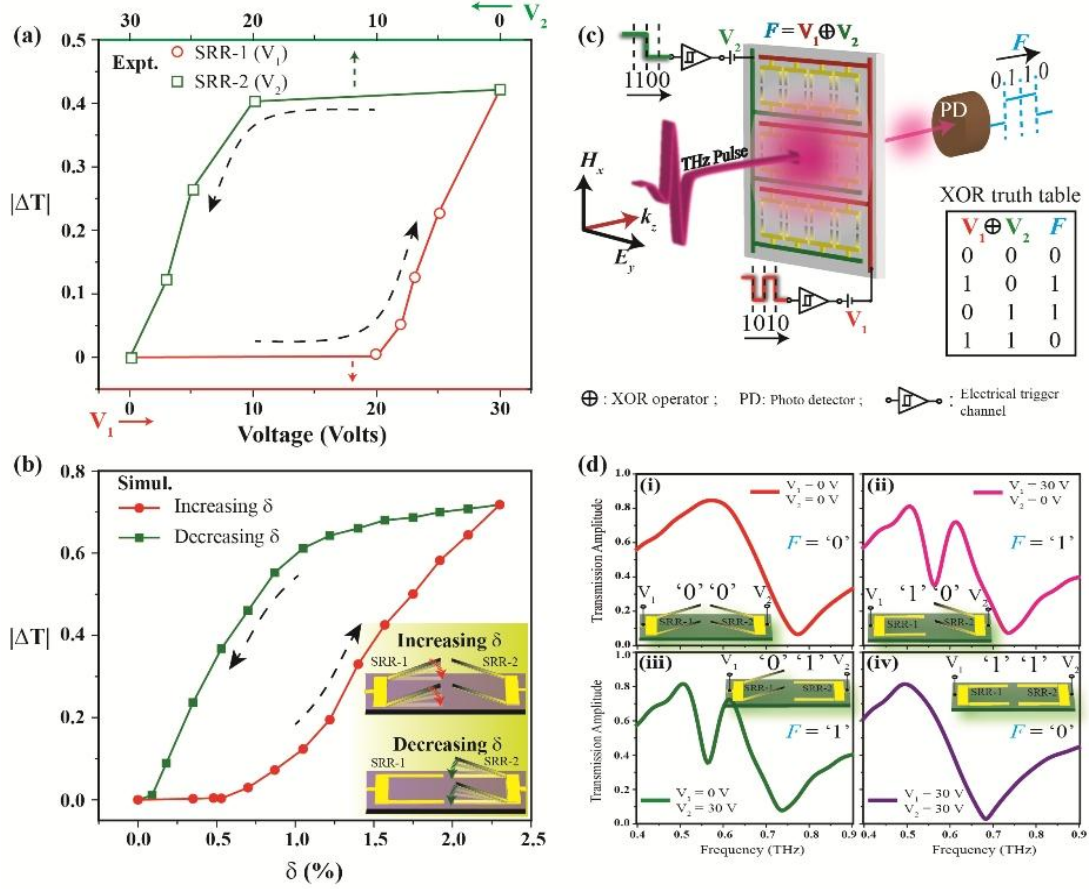

**Figure 3.** (**a**) Plot showing the measured Fano resonance transmission intensity ($|\Delta T|$) calculated for the curves shown in Fig. 2(a) and 2(b) respectively. The red circles represent increasing order of Fano resonance strength by applying $V_1$ on SRR-1 with $V_2 = 0$ V, whereas green squares represent decreasing state of Fano resonance in the presence of $V_2$ on SRR-2 with $V_1 = 30$ V. (**b**) Simulated Fano resonance intensity ($|\Delta T|$) showing two intensity states for a single asymmetry value ($\delta$) of the system. The inset figure represents the sequential actuation of SRR-1 and SRR-2 that governs the observed multiple-input-output (MIO) states for the MEMS Fano-metamaterial. (**c**) Pictorial representation of realizing the exclusive OR (XOR) logical operation (truth table) using the electrically controlled metastable states of resonators SRR-1 and SRR-2 determined by the respective voltage states $V_1$ and $V_2$ and an optical readout state (*F*) in the form of Fano excitation in the MEMS Fano-metamaterial. (**d**) Depict the measured THz spectrum for the various logical structural/voltage states of the MEMS Fano-metamaterial. (**i**) and (**iv**) show the symmetric configuration of the structures (0,0 and 1,1) that signifies the absence of Fano resonance (*F* = 0). (**ii**) and (**iii**) represents the asymmetric configuration of the structures (1,0 and 0,1) that results in the excitation of Fano resonance feature (*F* = 1) in the sample.

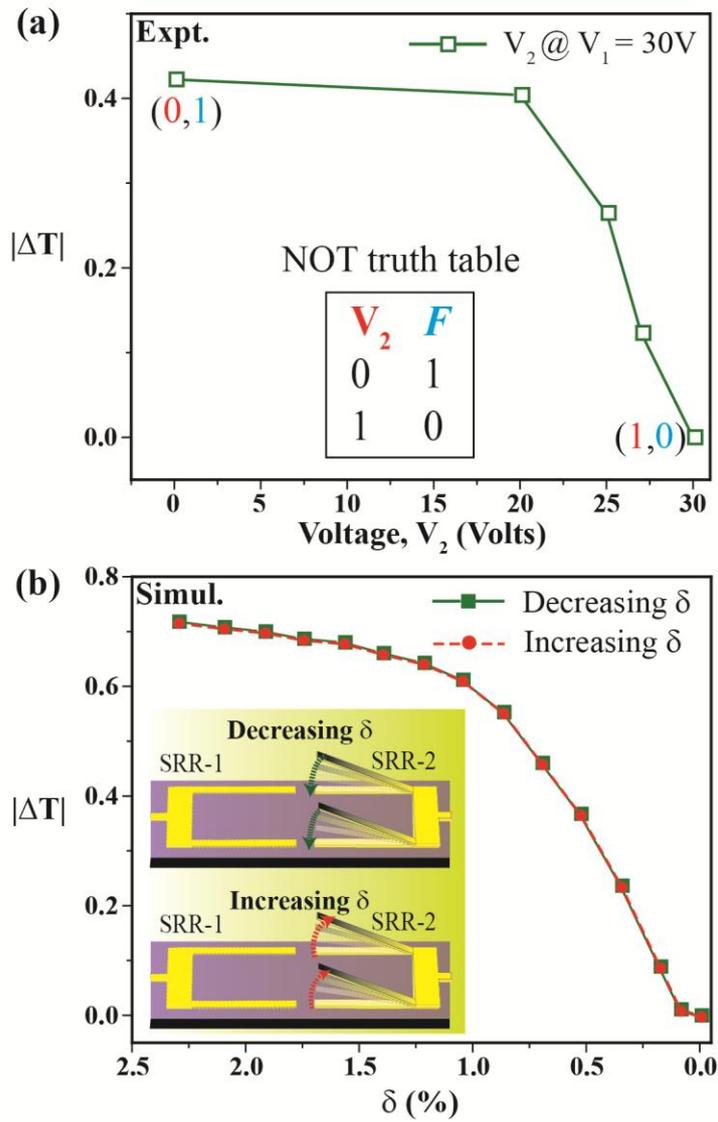

**Figure 4**. (**a**) Variation in |ΔT| of Fano resonance for increasing V2 for SRR-2, while keeping SRR-1 fixed on the substrate with the applied voltage $V_1$ = 30 V. (**b**) Variation in |ΔT| of Fano resonance for increasing and decreasing δ pathways, showing the single-input-output (SIO) configuration of the MEMS-Fano MM. The observed SIO configuration benefits in realizing the NOT logical operation in the form of true (1) and false (0) states of Fano resonance.

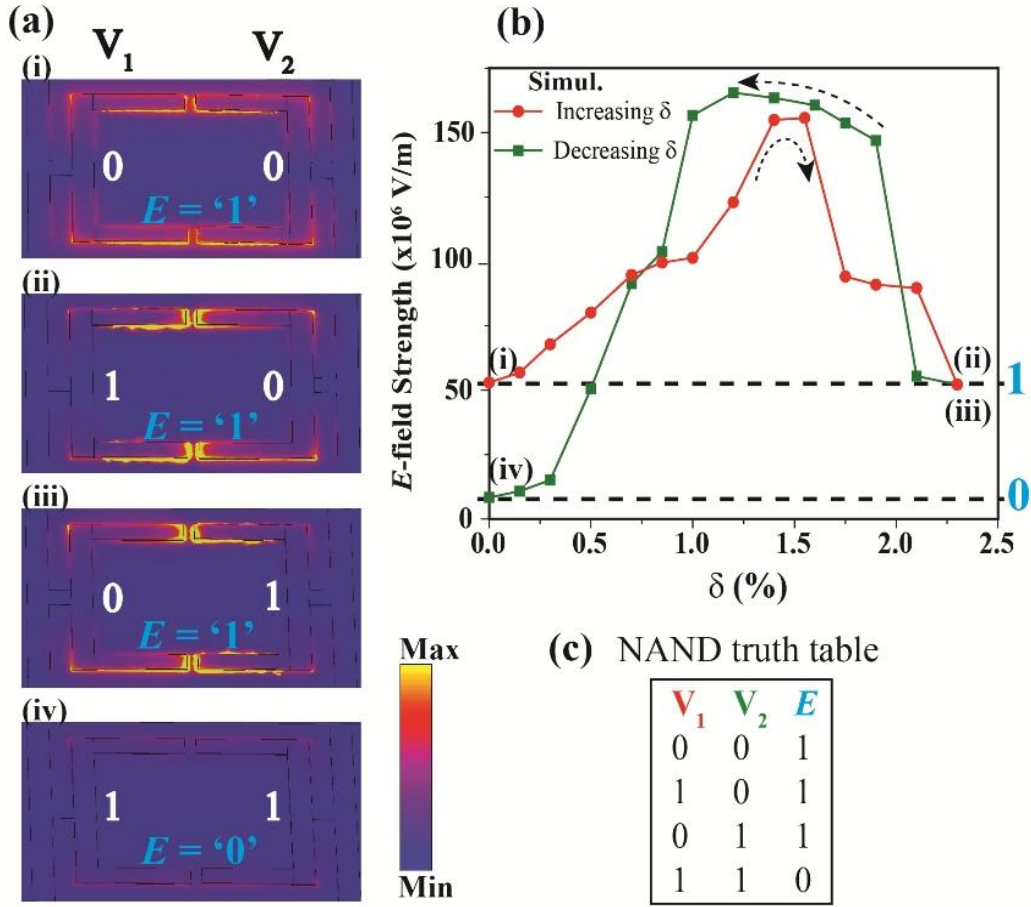

**Figure 5**. (**a**) The numerically calculated electric field distributions at 0.55 THz, where (**i**), (**ii**) and (**iii**) represent the true states of the electric field confinement (*E* = **1**), whereas (**iv**) shows the false state of *E* (i.e. *E* = 0), signifying the construction of NAND logical operation that is tabulated in (**c**). The input logical states 1 and 0 for $V_1$ and $V_2$ represent the 'up' and 'down' actuations states of SRR microcantilevers, respectively. (**b**) Distinctive variations shown for the enhanced spatially confined electric near-field strengths in the device at Fano frequencies (0.55 THz) that possess MIO behavior during sequential actuation of the SRR-1 (increasing δ) and SRR-2 (decreasing δ). The electric field strength is measured at the tip position of the SRR micro-cantilevers.